\documentclass[11pt,a4wide]{article}

\usepackage{amsfonts}
\usepackage{amsbsy}
\usepackage{epsfig}
\usepackage{latexsym}
\input amssym.def
\input amssym.tex
\def\a{{\mathcal{M}}}
\advance \topmargin by -\headheight
\advance \topmargin by -\headsep
\evensidemargin \oddsidemargin
\advance\hoffset by -3mm  
\advance\voffset by  8mm  
\def\bbbc{{\mathchoice {\setbox0=\hbox{$\displaystyle\rm C$}\hbox{\hbox
to0pt{\kern0.4\wd0\vrule height0.9\ht0\hss}\box0}}
{\setbox0=\hbox{$\textstyle\rm C$}\hbox{\hbox
to0pt{\kern0.4\wd0\vrule height0.9\ht0\hss}\box0}}
{\setbox0=\hbox{$\scriptstyle\rm C$}\hbox{\hbox
to0pt{\kern0.4\wd0\vrule height0.9\ht0\hss}\box0}}
{\setbox0=\hbox{$\scriptscriptstyle\rm C$}\hbox{\hbox
to0pt{\kern0.4\wd0\vrule height0.9\ht0\hss}\box0}}}}

\makeatletter
\@addtoreset{equation}{section}
\makeatother

\begin{document}
\hfuzz=100pt
\title{{\Large \bf{Schwarzschild-anti de Sitter
      within an Isothermal Cavity: Thermodynamics, Phase
      Transitions and the Dirichlet Problem}}} \author{\\M M
      Akbar\footnote{E-mail:M.M.Akbar@damtp.cam.ac.uk} \\ Department of
      Applied Mathematics and Theoretical Physics, \\ Centre for
      Mathematical Sciences, \\ University of Cambridge, \\
      Wilberforce Road, \\ Cambridge CB3 0WA, \\ U.K.}  \date{\today}
      \maketitle
\begin{center}
DAMTP-2003-17
\end{center}
\begin{abstract}
The thermodynamics of Schwarzschild black holes within an isothermal
cavity and the associated Euclidean Dirichlet boundary-value problem
are studied for four and higher dimensions in anti-de Sitter (AdS)
space.  For such boundary conditions classically there always exists a
unique hot AdS solution and two or no Schwarzschild-AdS black-hole
solutions depending on whether or not the temperature of the
cavity-wall is above a minimum value, the latter being a function of
the radius of the cavity.  Assuming the standard area-law of
black-hole entropy, it was known that larger and smaller holes have
positive and negative specific heats and hence are locally
thermodynamically stable and unstable respectively. In this paper we
present the first derivation of this by showing that the standard area
law of black-hole entropy holds in the semi-classical approximation of
the Euclidean path integral for such boundary conditions. We further
show that for wall-temperatures above a critical value a phase
transition takes hot AdS to the larger Schwarzschild-AdS within the
cavity. The larger hole thus can be globally thermodynamically stable
above this temperature. The phase transition can occur for a cavity of
arbitrary radius above a (corresponding) critical temperature. In the
infinite cavity limit this picture reduces to that considered by
Hawking and Page. The case of five dimensions is found to be rather
special since exact analytic expressions can be obtained for the
masses of the two holes as functions of cavity radius and temperature
thus solving exactly the Euclidean Dirichlet problem.  This makes it
possible to compute the on-shell Euclidean action as functions of them
from which other quantities of interest can be evaluated exactly. In
particular, we obtained the minimum temperature (for the holes to
exist classically) and the critical temperature (for phase transition)
as functions of the cavity-radius for five dimensions.
\end{abstract}

\section{Introduction}
Following the proposal of the AdS/CFT correspondence \cite{Maldacena}
which connects string theory on asymptotically anti-de Sitter space to
certain conformal field theories on its boundary, black holes in AdS
space have received renewed attention. In particular, the Hawking-Page
phase transition from hot anti-de Sitter (AdS) to Schwarzschild-AdS
\cite{HP} has been shown by Witten \cite{Witt1,Witt2} to correspond to
a phase transition from the confining phase to the deconfining phase
in the large $N$ limit of ${\cal N} = 4$ Yang-Mills on the boundary of
the AdS space. He argued that the strength of the conjecture is
demonstrated by the existence of a holographic duality even at
non-zero temperature where supersymmetry and conformal invariance are
broken.

In the Euclidean picture the boundary at which the conformal field
theory is defined is an $S^{1}\times S^{n-1}$ which is a
codimension-one slice of the Euclideanised AdS and Schwarzschild-AdS
at radial infinity. However, one can consider a more general, and
non-trivial, boundary by taking the $S^{1}\times S^{n-1}$ at a finite
radial distance so that the $S^{n-1}$ is of finite volume (and
assuming, initially, that the $S^1$-fibre has finite radius as
well). Infinite volume for either or both of $S^1$ and $S^{n-1}$ occur
as limits. From the holographic point of view, such a finite boundary
is relevant for understanding the stronger form of holography. Indeed
a large literature exists where this has been considered in the
context of the Randall-Saundrum type scenarios \cite{RS} and various
related topics (see \cite{Nojiri} for a review and references
therein).
A primary objective in such studies as in the infinite boundary
limit mentioned above, is to understand the relationship between
the geometries of the possible regular bulk solutions and the
geometry of the given boundary. Such study falls under the purview
of the classical Dirichlet boundary-value problem in Riemannian
geometry (see below).  Once the relationship between various bulk
geometries and the boundary geometry are known, either
qualitatively or quantitatively, one can explore the
semi-classical implications using Euclidean path integral
formulation. In particular, if the geometries of the bulk are
known as functions of the boundary variables, one can find the
classical action and other quantities of interest purely in terms
of boundary data. In the example above, in which the boundary is
at infinity, this can be done precisely for arbitrary dimensions.
The meaningful boundary data then is the ratio of the two radii of
the fibre and the base. However, when the boundary is brought at a
finite distance this becomes rather non-trivial as we will see in
this paper.

In the path integral formulation of Euclidean quantum gravity finite
$S^{1}\times S^{n-1}$ boundaries first appeared in the study of black
holes in a ``box''. The radii $\alpha$ and $\beta$ of the
$S^{1}$-fibre and $S^{n-1}$-base (with the standard metrics on them)
constitute the canonical boundary condition with the interpretation
that the $S^{n-1}$ cavity of radius $\alpha$ is immersed in an
isothermal heat bath of temperature $T=\frac{1}{2\pi\beta}$.  It was
shown in \cite{York1} that within a finite $S^2$ cavity apart from hot
flat space, classically there are two or no black-hole solutions
depending on whether the temperature of the heat bath is above or
below a certain value\footnote{Least the terminology be a source of
confusion due to their connotations, the ``cavity'' here is the
$S^{n-1}$ base and is not the same as what we called the ``boundary''
in the Euclideanised picture which is an $S^{1}\times S^{n-1}$.}. This
is true for any finite radius of the cavity. Of the two solutions, the
larger one has a positive specific heat and the smaller one a negative
specific heat. Hence, only the larger solution is locally
thermodynamically stable. It is possible to find the masses of the two
black holes as functions of the cavity radius and wall-temperature
(i.e., in terms of $\alpha$ and $\beta$ in the Euclidean picture). For
sufficiently high temperature the larger mass solution has a more
negative action than hot flat space and hence a stable black hole can
nucleate from hot flat space in a thermodynamically consistent manner
within a cavity of arbitrary radius. In the infinite-volume limit of
the cavity only the smaller-mass solution with the negative specific
heat survives. Therefore, only in a finite cavity is it possible to
have a thermodynamically stable black hole solution for such boundary
conditions. Recently these results have been generalized and discussed
for higher dimensions in \cite {AG}. However, in the case of a
negative cosmological constant, stable Schwarzschild-AdS$_{n+1}$ black
holes are possible \cite{HP} and one is not required to introduce an
isothermal finite cavity, at least for obtaining a stable solution.
This perhaps explains the comparative lack of attention in the
literature to the finite cavity case for $\Lambda <0$ except the study
made in \cite {BCM} where this has been studied for four
dimensions. There it is shown that for such boundary condition, there
are two/no Schwarzschild-AdS$_4$ solutions and always a AdS$_4$
solution. Assuming that the standard area-law of black-hole entropy
holds for such boundary conditions, there it was shown that the larger
and the smaller black hole have positive and negative specific heats
respectively as in the $\Lambda=0$ case.  As the cosmological constant
introduces a scale, it is not possible to find the masses in terms of
the boundary variables, which was possible in the flat-space case for
four dimensions \cite{York1}, except in the limit of infinite
cavity. In this paper we will be extending this picture to higher
dimensions by providing a complete semi-classical treatment. In the
classical side of the picture we will discuss solutions of bulk
geometries in terms of boundary data, i.e., the two radii
$(\alpha,\beta)$ and will combine this with the semi-classical study
to show that a phase transition occurs within the cavity for
sufficiently high value of its wall-temperature. In the infinite
cavity limit the analogue of the phase transition is the well-known
Hawking-Page phase transition, in which case the phase transition is
studied as functions of the Hawking temperature (or equivalently the
horizon radius) of the black holes instead of the boundary variables
(which is the only way to do this since the local temperature of any
finite Schwarzschild-AdS redshifts to zero at infinity, unlike in the
flat space case, see later). Such a study has not been done before for
$\Lambda<0$ for four or higher dimensions.

The goal of this paper is therefore twofold: to study possible black
hole solutions as functions of the boundary geometry and to compute
their on-shell actions, and explore which solutions dominate and under
what conditions on the boundary variables. We will be as rigorous as
possible in our treatment and obtain explicit formulae and relate
them to the well known $\Lambda=0$ results.  Interestingly, we find
that a precise quantitative treatment is possible only in five
dimensions. This is the most interesting because of its connection
with the 4-D world within a holographic context. The results obtained
herein can thus be extended in various directions and sets the ground
for many future investigations.

This paper is arranged in the following way. In Section 2 we
discuss the Euclidean AdS$_{n+1}$ and Schwarzschild-AdS$_{n+1}$
metrics and set our conventions. In Section 3 we discuss the
Dirichlet problem and thermodynamics of the black-hole solutions
after briefly discussing the $\Lambda=0$ case. Phase transitions
are studied in Section 4. In Section 5 we specialize to five
dimensions and show how the above questions can be addressed
precisely. We conclude with remarks on possible ramifications of
the results for current topical issues.

\section{Euclidean AdS$_{n+1}$ and Schwarzschild-AdS$_{n+1}$}
The Euclidean AdS$_{n+1}$ and Schwarzschild-AdS$_{n+1}$ metrics have
the following forms:
\begin{equation}
ds^{2}=V(r)\,
dt^{2}+V(r)^{-1}dr^{2}+r^{2}d\Omega_{n-1}^{2}. \label{shads}
\end{equation}
with
\begin{equation}
V(r)=\left(1+\frac{r^{2}}{l^{2}}\right),
\end{equation}
and
\begin{equation}
V(r)=\left(1-\frac{\mu}{r^{n-2}}+\frac{r^{2}}{l^{2}}\right),\label{shads1}
\end{equation}
respectively. Both metrics satisfy the Einstein equation with a
cosmological constant $\Lambda=-\frac{n(n-1)}{2l^2}$. The quantity
$\mu$ in Eq.(\ref{shads1}) gives the mass $m$ of the black hole
\cite{Myers}
\begin{equation}
\mu=\frac{16\pi G\,m}{(n-1) \mathrm{Vol}(S^{n-1})}.
\end{equation}
For notational convenience we will set $l=1$ without any loss of
generality and denote $r$ as $\rho$ and $\mu$ as $M$ in this
scale. When there is no chance of confusion we will refer to $M$ as
the ``mass'' of the black hole.  The horizon of the
Schwarzschild-AdS$_{n+1}$ metric (\ref{shads}) is the positive root
of the equation
\begin{equation}
V(\rho)\equiv \left(1-\frac{M}{\rho^{n-2}}+\rho^{2}\right)=0 \label{hor}
\end{equation}
and will be denoted by $\rho_{b}$. The horizon radius $\rho_{b}$ and the
mass $M$ are in 1-1 correspondence:
\begin{equation}
M=\rho^{n}_b+\rho^{n-2}_b.\label{massads}
\end{equation}
The singularity at the horizon can be removed if we give $t$ a
periodicity of
\begin{equation}
\Delta{t}=\frac{2\pi}{\kappa}\label{per}
\end{equation}
where
\begin{equation}
\kappa \equiv
\frac{1}{2}\,V'(\rho_{_b})=\frac{(n-2)}{2}\frac{M}{\rho_{_b}^{n-1}}+\rho_{_b}=
\frac{1}{2}\frac{n\,\rho_{_b}^2+n-2}{\rho_{_b}}
\label{perid}
\end{equation}
is the surface gravity. The metric is then well-defined for
$\rho_{_b} \le \rho < \infty$ which includes the horizon which is
now the $(n-1)$-dimensional fixed point set of the Killing vector
$\partial/\partial \tau$ and is a regular bolt of the metric
(which explains the use of subscript in $\rho_{_b}$). The
periodicity $ \Delta{t}$ gives the inverse of Hawking temperature
of the hole
\begin{equation}
T_{H}=\frac{1}{4\pi}\frac{n\,\rho^2_{_b}+n-2}{\rho_{_b}} \label{Hawk}.
\end{equation}
Note that the AdS$_{n+1}$ metric is regular for $0\le r \le
\infty$. One therefore is not required to ascribe a certain
periodicity to $t$ in order to achieve regularity. In other words, one
can choose the periodicity of $t$ arbitrarily as in the case of flat
space.
\section{Black holes in an isothermal cavity and the Dirichlet problem}
As mentioned in the introduction, the problem that we address in this
paper formally falls under the scope of the classical Dirichlet
problem in which one seeks one or more regular $(n+1)$-dimensional
Riemannian solutions $({\cal{M}},g_{\mu\nu})$ of the Einstein
equations for a given $n$-boundary $(\Sigma,h_{ij})$ such that
$\partial{\cal{M}}=\Sigma$ and
$g_{\mu\nu}|_{\partial{\cal{M}}}=h_{ij}$ possibly with the condition
of regularity. In most physically interesting cases, the Dirichlet
problem simplifies as only cohomogeneity-one metrics whose principal
orbits share the topology and symmetry of the boundary are
considered. In the presence of only a possible cosmological constant
term, this reduces the Einstein equation to a set of ordinary
differential equations to be solved subject to the boundary condition
and regularity. However, in cases of high symmetry the general
solution of this set and the manifolds over which they can be extended
completely or partially may be known in advance. The Dirichlet problem
then simplifies to the problem of embedding $\Sigma$ in known
manifolds -- the ``infilling'' solutions then are the compact regular
parts of the manifolds enclosed by $\Sigma$. Such solutions provide
semi-classical approximations to the path integral and are the
starting point of a quantum treatment. Depending on the boundary data
there can be zero, one or multiple infilling solutions of similar or
different topologies even when one assumes a high degree of symmetry
for the codimension-one slices.

Because the classical action consists of a bulk and boundary term, if
the geometries of the possible bulk solutions are known as functions
of the boundary variables one can find the on-shell Euclidean actions
purely in terms of the boundary variables.  Knowledge of the bulk
geometries in relations to the boundary geometry is all that one needs
to obtain for semi-classical considerations. However, it may not be
possible to obtain the infilling geometries as analytic functions of
the boundary data which then requires one to adopt an indirect
approach.

For either AdS$_{n+1}$ or Schwarzschild-AdS$_{n+1}$, with the
periodic identification of the $t$-coordinate, a hypersurface at a
constant value of the radial coordinate $\rho$, say $\rho=\rho_{0}$,
has the trivial product topology $S^1\times S^{n-1}$ and is endowed
with the $n$-metric $h_{ij}$ given by
\begin{equation}
ds_{\Sigma}^2=\beta^2\,dt^2+ \alpha^2 d\Omega_{n-1}^{2}. \label{buonmet}
\end{equation}
If one excises the part of the manifold for which $\rho>\rho_{0}$
one is left with a non-singular compact manifold with a boundary
which, by construction, provides an infilling solution for
$(\Sigma,h_{ij})$. Therefore for a given $S^1\times S^{n-1}$
boundary there are two topologically distinct possibilities for
the infillings.

Comparing Eqs (\ref{buonmet}) and (\ref{shads}), one obtains
$\rho_{0}=\alpha$ trivially for either AdS$_{n+1}$ or
Schwarzschild-AdS$_{n+1}$. For a given $S^1\times
S^{n-1}$ boundary the periodicity of the $t$ coordinate of the
AdS$_{n+1}$ metric is given by
\begin{equation}
\sqrt{1+\alpha^{2}}\,\,\Delta t =2\pi\,\beta.
\label{dirads00}
\end{equation}
Obviously for a given set $(\alpha,\beta)$ the periodicity is
unique\footnote{This slightly different convention for notation, which
makes $\beta$ the ``radius'' of the $S^1$ fibre, would avoid the
recurrent appearance of $\pi$ in the rest of the paper.}.  On the
other hand, for Schwarzschild-AdS$_{n+1}$ one needs to find the
periodicity via (\ref{per}) by solving the following equation for $M$
first:
\begin{equation}
\sqrt{1-\frac{M}{\alpha^{n-2}}+\alpha^{2}}=\beta\, \kappa(M). \label{mast}
\end{equation}
This is the standard relationship between the local temperature at the
cavity wall and the Hawking temperature
\begin{equation}
T_{loc}=\frac{T_{H}} {\sqrt{g_{00}}} \equiv \frac{T_{H}}{
\sqrt{\left(1-\frac{M}{\alpha^{n-2}}+\alpha^{2}\right)}}.
\end{equation}
Eq.(\ref{mast}) leads to a complicated algebraic expression in
$M$. Obviously there is no {\it{a priori}} obstruction as one can {\it
start} with a Schwarzschild-AdS with known mass and take a constant
$\rho$-slice for which Eq.(\ref{mast}) is true trivially. However, it
is not clear for what values of $\alpha$ and $\beta$ black-hole
solutions exist and what their possible number and properties are.  We
will be addressing these questions in the following sections. But
before doing that we briefly recapitulate what we know about the
$\Lambda=0$ case which is a limiting case of the present study
(i.e. $\l\rightarrow\infty$ limit).  The results for $\Lambda=0$ will
be a constant reference for judging success of the $\Lambda<0$
case. The results for $\Lambda=0$ will also help us understand the
limit when the cavity radius is very small compared to $l^2$, i.e.,
when $\alpha<<1$.
\subsection{The case of zero cosmological constant}
For $\Lambda=0$, Eq.(\ref{mast}) simplifies considerably and can be
rewritten as \cite{AG}
\begin{equation}
x^{n}-x^{2}+ p=0 \label{mast0},
\end{equation}
where $x \equiv M^{\frac{1}{n-2}}/\alpha$ and $p
=\frac{(n-2)^{2}}{4}\frac{\beta^{2}}{\alpha^{2}}$.  It is easy to see
that there are in general two solutions if
\begin{equation}
p \le \left(\frac{2}{n}\right)^\frac{2}{n-2}\left(1 -\frac{2}{n}\right).\label{requireads}
\end{equation}
For a given $\alpha$, this sets an upper limit to $\beta$, i.e., this
gives a lower limit to the temperature of the cavity-wall
\begin{equation}
T \ge
T_m\equiv \frac{1}{4\pi}\left(\frac{n}{2}\right)^{\frac{1}{n-2}}\sqrt{n(n-2)}
\,\frac{1}{\alpha}. \label{flatmin}
\end{equation}
For $T<T_m$ no black-hole solution exists. For $T=T_m$ the two solutions are
degenerate and is given by
\begin{equation}
\rho_b=\left(\frac{2}{n}\right)^{\frac{1}{n-2}}\alpha.\label{massFLAT}
\end{equation}
For $T>T_m$, there are always two positive roots of
Eq.(\ref{mast0}). For a given cavity, the higher the
wall-temperature the heavier (lighter) is the larger (smaller)
solution.

For four dimensions the Eq.(\ref{mast0}) is cubic and is solvable
\cite{York1}. However, apart from a few other special values of $n$
this is not solvable using ordinary algebraic methods. Fortunately,
Eq.(\ref{mast0}) falls under a special class of algebraic equations --
commonly known as trinomial equations in the mathematical literature
-- which can be solved using higher order hypergeometric functions in
one variable. For arbitrary $n$ and $p$ the solutions have been
obtained in \cite{AG}. Thus the corresponding Riemannian Dirichlet
problem is exactly solvable for arbitrary
dimension\footnote{Restricting of course within the class of
cohomogeneity one metrics whose principal orbits share the symmetry
and topology of the boundary.}.

\subsection{$\Lambda<0$: Schwarzschild-AdS$_{n+1}$ solutions}
For $\Lambda=0$ the problem reduces to studying Eq.(\ref{mast0}) for
which it is only the squashing $\alpha/\beta$ that matters. For
$\Lambda<0$, the analogue of Eq.(\ref{mast0}) reads
\begin{equation}
F:={\rho}^{n+2}_{_b}+{\rho}^{n}_{_b}+\frac{1}{4}{\alpha}^{n-2}{\beta}^{2
}{n}^{2}{\rho}^{4}_{_b}+\frac{1}{2}{\alpha}^{n-2}\left
[{\beta}^{2}n(n-2)-2\,({ \alpha}^{2}+1)\right
]{\rho}^{2}_{_b}+\frac{1}{4}{\alpha}^{n-2}{\beta}^{2}\left (n-2\right
)^{2}=0 \label{eq}.
\end{equation}

This is obtained from Eq.(\ref{mast}) by simply eliminating the
square-root and using Eq.(\ref{massads}).  Since $M$ and $\rho_{_b}$
are in 1-1 correspondence both are in principle equivalent
variables. However, use of $\rho_{_b}$ leads to algebraic
simplifications, albeit the resulting equation is still difficult to
solve exactly.

Only the positive roots $\rho_{_{b}}<\alpha$ of Eq.(\ref{eq}) will
give the mass and hence the geometries of the infilling black
holes. Note that Eq.(\ref{eq}) cannot be reduced to a one-parameter
problem by any redefinition of variables. This is true even for the
infinite cavity limit although a different kind of simplification
occurs in this limit. The infinite cavity limit has been studied
rigorously in \cite{HP, Witt1,Witt2} and will occur as a special case
in our study.

In four dimensions Eq.(\ref{eq}) is quintic and hence, ordinary
algebraic methods fail to produce analytic solutions for the masses of
the two black holes for four dimensions in terms of radicals unlike
the $\Lambda=0$ case found in \cite{York1}. One can, however, solve
Eq.(\ref{eq}) in terms of hypergeometric functions by following
Birkeland's general solution of algebraic equations of order $n$ with
arbitrary coefficients \cite{Birkeland} or by using
${\cal{A}}$-hypergeometric functions \cite{Stu}. Unlike the
$\Lambda=0$ case, the solutions will be in terms of hypergeometric
functions of {\it{several}} variables. The variables will be
non-trivial functions of $\alpha$ and $\beta$.  This method would be
unsuitable for obtaining direct information as the cavity radius and
temperature are varied and hence will not be used in the rest of the
paper. We will return to the issue of explicit solutions in five
dimensions in Section 5.  However, as we will see below it is possible
to study the infilling solutions and their thermodynamics without
requiring explicit solutions.

\subsection{Number of solutions}
As mentioned earlier, the only study for finite cavity was made in
\cite{BCM} for four spacetime dimensions. It was shown there that
there are two black hole solutions with negative and positive specific
heats within the cavity. Both survive in the infinite cavity-radius
limit unlike the Schwarzschild case in flat space. We will now see that
this is true for higher dimensions as well\footnote{For arbitrary
dimensions the only mention of a finite cavity appears briefly by
Prestidge in \cite{Tim} who ultimately considers the infinite volume
limit to search for the negative modes predicted in \cite{HP}.}.

It is easy to see that Eq.(\ref{eq}) does not admit positive solutions
if $\beta^{2}\ge\frac{2(\alpha^2+1)}{n(n-2)}$ since no changes of sign
occur in the coefficients of the various powers which are all positive
in this case. Only the coefficient of $\rho^2_{_b}$ can be negative
and this happens if
\begin{equation}
\beta^{2}<\frac{2(\alpha^2+1)}{n(n-2)}\label{sign2}.
\end{equation}
There can be up to two positive roots and up to two (for $n$ even) or
three (for $n$ odd) negative roots. Note that (\ref{sign2}) places a
necessary, and not a sufficient condition on $\beta$ (equivalently
$T$). For $n$ even, the quantity $(-1)^{n+2}{\beta}^{2}\left
(n-2\right )^{2}$ is positive and hence there can never be one
positive root and one negative root. The possibility of having a
single positive root is thus ruled out (a double-root is counted
twice). Since complex roots appear in pairs, similar arguments apply
for $n$ odd. Thus, {\it {there will be two positive roots (double root
counted twice) or no positive roots}}.  Only the positive roots which
are less then the cavity radius $\alpha$ can qualify as black-hole
solutions inside the cavity. It is easy to check that this would
automatically be the case for the two possible positive roots of
Eq.(\ref{eq}). To see this rewrite Eq.(\ref{eq}) in the following
form:
\begin{equation}
{\beta}^{2}={\frac
{4\,\rho_{_b}^2\left({\alpha}^{n}+\,{\alpha}^{n-2}-\rho^n_{_{b}}-{\rho^{n-2}_{_b}}\right)}{{\alpha}^{n-2}\left
  ({n}\,\rho^2_{_b}+n-2\right)^2}}.
\label{betaasrho}
\end{equation}
$\beta^2$ is a single-valued function of $\rho_{_b}$ and is positive
if and only if $\rho_{_b}\in (0,\alpha)$. It is continuous and
well-defined within this interval (the denominator is strictly
positive). This is true for $n\ge3$ and for any positive value of
$\alpha$. Since the maximum number of positive roots of Eq.(\ref{eq})
is two, it immediately follows that $\beta^2$ grows from zero (at
$\rho_{_b}=0$) and ends with zero (at $\rho=\alpha$) with {\it{one}}
``hump'' in the middle, i.e., $\beta^2$ increases monotonically to a
maximum $\beta^2_{m}$ and then decreases monotonically to zero (Fig.
5.1). Therefore there will be two $\rho_{_b}$, i.e., two black-hole
solutions for a given boundary with specified $(\alpha,\beta)$
provided $\beta$ (or $T$) is equal to or less (greater) than the
minimum value needed for the cavity radius of $\alpha$. Because
$\beta$ is a continuous function of $\rho_{_b}$ within the interval
$(0,\alpha)$ the two black-hole solutions exist for any temperature
above the minimum value without any discontinuity. This holds for
arbitrary cavity-radius. The minimum temperature needed for the
solutions to exist is a function of the radius of the cavity and will
be discussed in Section 5.3.4.
\begin{figure}[!h]
  \begin{center}
    \leavevmode
    \vbox {
      \includegraphics[width=10cm,height=6cm]{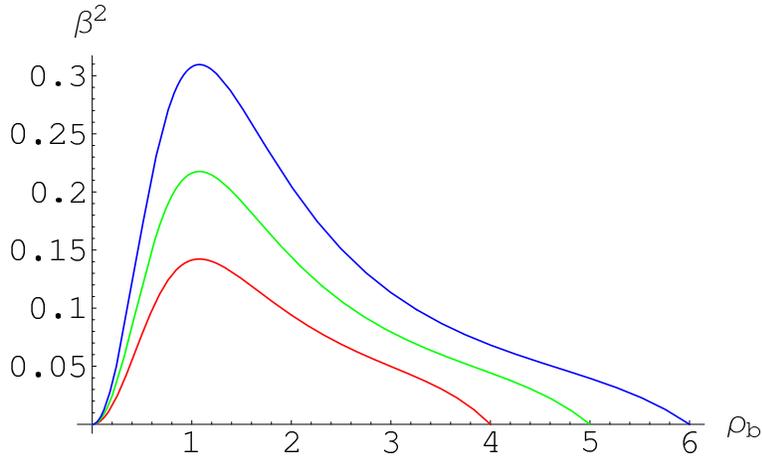}
      \vskip 3mm
      }
    \caption{$\beta^2$ (for $n=11$) as a function of $\rho_{b}$ for
    $\alpha=4, 5 ,6$ :
    there is a unique maximum in each case. Note that the curves
    corresponding to different values of $\alpha$ do not cross as
    explained in Section $3.2$.}
    \label{Fig1}
  \end{center}
\end{figure}
Note that the shape of the curve for $\beta^2$ as a function of
$\rho_{_b}$ is not obvious from Eq.(\ref{betaasrho}) alone
because of the non-triviality of the denominator. The fact that
there will be two/no positive roots, as we have shown before, is
crucial to the argument for the general number of solutions.
\subsection{The minimum temperature $T_m$ as a function of cavity radius}
For a cavity of fixed radius the minimum temperature $T_{m}$ required
for the black-hole solutions to exist is the one corresponding to the
maximum of $\beta\equiv \beta_{m}$ at which the solutions are
degenerate. By differentiating Eq.(\ref{betaasrho}) one obtains that
at $\beta= \beta_{m}$
\begin{equation}
\begin{array}{rcl}
n(n-2)\rho^{n+2}_{_{b_m}}+2\left (n^2-2{n}-2 \right
)\rho_{_{b_m}}^n+n(n-2)\rho_{_{b_m}}^{n-2}\!\!\!\!&+\!\!\!\!&2{n}\left
({\alpha}^{n}+{\alpha}^{n-2}\right ){\rho^2_{_{b_m}}}\\
\!\!\!\!&-\!\!\!\!&2\left (n-2\right )\left
({\alpha}^{n}+{\alpha}^{n-2} \right )=0\label{minmin}
\end{array}
\end{equation}
where $\rho_{_{b_m}}$ is the value corresponding to the maximum (minimum) of $\beta$ (\,$T$) (and does not stand for the
minimum/maximum value of $\rho_{_b}$).  One obtains the minimum
temperature as a function of the cavity radius by directly substituting the
root of Eq.(\ref{minmin}) in Eq.(\ref{betaasrho})\footnote{One can
check that Eq.(\ref{minmin}) has a positive root in the interval
$(0,\alpha)$. However, this trivially follows from our previous
analysis.}. However, note that, like
Eq.(\ref{eq}), Eq.(\ref{minmin}) is not generally solvable using
ordinary algebraic methods.

For $n\ge 3$, all terms in Eq.(\ref{minmin}) are positive except the
last. It therefore follows that
\begin{equation}
2n(\alpha^n+\alpha^{n-2})\rho_{_{b_m}}^2-2(n-2)(\alpha^n+\alpha^{n-2})<0 \label{upp}
\end{equation}
implying
\begin{equation}
\rho_{_{b_m}}<\sqrt {{\frac {n-2}{n}}}. \label{uppbon}
\end{equation}
After dividing Eq.(\ref{minmin}) by $(\alpha^n+\alpha^{n-2})$ it is
easy to see that the larger the value of $\alpha$ the larger is
$\rho_{b_m}$. This accounts for the gradual shift of the peak to the
right in Fig. 5.1 for larger values of cavity radius. At the infinite
cavity limit (\ref{upp}) is replaced by an equality giving
\begin{equation}
\rho_{_{b_m}}=\sqrt {{\frac {n-2}{n}}}\label{uppbonn}
\end{equation}
exactly. This therefore gives an absolute upper bound on the horizon
radius, and hence the entropy, of the smaller black hole irrespective
of the cavity radius and temperature.  Note that Eq.(\ref{uppbonn})
can be obtained by differentiating the Hawking temperature given
by the inverse of $\Delta t$.  With a finite cavity, one needs to vary
the product of the Hawking temperature multiplied by the Tolman red-shift
factor for which the extremum occurs at a different value of
$\rho_{_b}$. For finite cavity this value of $\rho_{_b}$ is less than
that obtained by varying the Hawking temperature alone which is
Eq.(\ref{uppbonn}) and gradually increases asymptotically to
(\ref{uppbonn}) as one increases the cavity radius.

\subsection{Geometry of Eq.(\ref{eq})}
The difficulties in obtaining exact solutions of Eq.(\ref{minmin}) does
not preclude us from observing the following general fact
: $T_m (\alpha)$ increases with decreasing
radius of the cavity. This is because the curve for $\beta$ as a
function of $\rho_{_{b}}$ for any fixed value $\alpha$ of the cavity
radius completely covers the curve corresponding to a lower value of
$\alpha$. We have already noticed this from Fig. 5.1. To see that this
holds in general consider the converse: if two curves of $\beta$
(corresponding to two different values of cavity radius $\alpha$) as a
function of $\rho_b$ meet at some point in the $\beta^{2}-\rho_{_b}$
plane, the two values of $\alpha$ should satisfy Eq.(\ref{eq}) for the
same pair $(\beta^2,\rho_{_b})$. Treating Eq.(\ref{eq}) as an equation
for $\alpha$, it is easy to see that there can be only one/no positive
solution for $\alpha$ for a given $(\beta^2,\rho_{_b})$. Therefore
nowhere in the $\beta^{2}-\rho_{_b}$ plane can two curves
corresponding to two different values of $\alpha$ meet. The curves
corresponding to lower values of $\alpha$ will be within the envelopes
of those corresponding to higher values of $\alpha$. Therefore the
minimum temperature $T_m$ needed for the existence of two black-hole
solutions increases with decreasing cavity radius. Since
$\beta$ (as a function of $\rho_{_b}$) is continuous, for any
arbitrarily higher temperature the two solutions continue to exist.

As derivatives of $\beta^2$ with respect to $\alpha$ are smooth, the
curves in the $\beta^{2}-\rho_{_b}$ plane for various values of
$\alpha$ fill in densely. The two dimensional surface of
Eq.(\ref{eq}) in the $\alpha^2-\beta^{2}-\rho_{_b}$ space is therefore
smooth and continuous. Therefore $T_m$ is a smooth function of the
cavity radius and increases with decreasing value of the radius. It
remains to see whether it is possible to obtain $T_m$ as a function of the
cavity radius. We will return to this issue in section 5.1.

\subsection{Infinite Cavity Limit}
For any finite Schwarzschild-AdS black hole the local temperature at
infinity is zero. Therefore the above picture -- where one fixes the
cavity radius and its wall-temperature -- is not well-defined in the
infinite limit of the cavity in the case of Schwarzschild-AdS. One
cannot simply fix a non-zero temperature at infinity and look for a
finite Schwarzschild-AdS black-hole solution in the interior. This,
however, is only possible in the $\Lambda=0$ limit. For vanishing
temperature at infinity all Schwarzschild-AdS black holes are equally
good infilling solutions.

However, the product of the local temperature and the cavity radius is
well-defined in the infinite cavity limit\footnote{In the Euclidean
picture, it corresponds to the fact that the ratio of the radii of the
$S^1$ fibre and the $S^{n-1}$ base remains finite.}. This product
essentially is the Hawking temperature (\ref{Hawk}) and substitutes
the wall-temperature of the cavity in this limit. One can now study
the thermodynamics of black holes in terms of the Hawking temperature
alone. This has been done in \cite{HP}. For a given Hawking
temperature, there are in general two black holes -- this corresponds
to the doubled-valuedness of the former as one can see from
(\ref{Hawk}). The Hawking temperature has a minimum, $T_m=
\frac{1}{2\pi}\sqrt{n(n-2)}$ which corresponds to a horizon-radius of
$\rho_b=\sqrt{(n-2)/n}$ in conformity with our analysis above. The
mass of one hole decreases and the other increases as one raises the
Hawking temperature.

One can immediately see the simplifications arising in the
infinite limit of the cavity. It is rather trivial to find the
masses of the two black holes from the Hawking temperature
(\ref{Hawk}) in contrast to the finite-cavity case where one needs
to solve (\ref{eq}). Further simplification arises in this limit
when one considers semi-classical implications. This will appear
in the next section in which we will explore the semi-classical
implications for arbitrary finite cavity from which known results
in the infinite limit of the cavity will be reproduced as a
special case.

\section{On-shell Actions, Semi-classical Results and Phase Transition}
Our discussion so far was purely classical. However, it has set the
ground for semi-classical considerations as we will see below.  First,
let us briefly recall the facts for $\Lambda=0$ case. We have already
mentioned that the Dirichlet problem is exactly solvable in this
case. This enables one to find the classical actions $I_{E}$ of the
infilling black-hole solutions as functions of boundary variables and
thus study semi-classical physics. The specific heat of the larger
solution is positive and that of the smaller solution is
negative. From the action one also finds that the larger mass solution
in $(n+1)$-dimensions, which has a positive specific heat, has a lower
action than hot flat space, for temperatures \cite{AG}
\begin{equation}
T>T_c\equiv \frac{1}{4\pi}\frac{n-2}{\sqrt{\left ({\frac
{4(n-1)}{{n}^{2}}}\right )^\frac{2}{\left (n-2\right )}-\left ({\frac
{4(n-1)}{{n}^{2}}}\right )^\frac{n}{\left (n-2\right )}}}\,\frac{1}{\alpha}.
\end{equation}
The lower mass solution always has a larger action than flat
space. Thus above $T_c(\alpha)$ the larger black-hole solution can
spontaneously nucleate from hot flat space within the cavity in a
thermodynamically consistent manner in which the free energy of the
system, $ F= I_{E}/ (2\pi\beta)$, does not increase. As one takes the
cavity radius to infinity the larger solution fills in the entire
cavity irrespective of its wall temperature and hence becomes
irrelevant and only the smaller solution exists.  Therefore,
black-hole nucleation is thermodynamically consistent only within a
finite cavity.  Note that the critical temperature $T_c$ is inversely
proportional to the cavity radius and is higher than the minimum
temperature $T_m$ (\ref{flatmin}) needed for the two black hole
solutions to exist {\it classically}. We will see below that the
situation is similar for anti-de Sitter space as well. To the best of
our knowledge, such a study has not been carried out before for four
or higher dimensions except the study made in infinite cavity limit in
\cite{HP} and which will reappear as a special case below.
\subsection{Action of the Infilling Black Holes}
For an $(n+1)$-dimensional manifold ${\cal{M}}$ with an
$n$-dimensional boundary ${\partial \a}$, the Euclidean action is
\cite{York-sur,GHsur,Regge1, Regge2}:
\begin{equation}
I_{E} = -\frac{1}{16\pi G}\int_{\a} d^{n+1} x \sqrt{g} \left(R
-2\Lambda\right)- \frac{1}{8\pi G}\int_{\partial \a} d^n x \sqrt{h}\,
K \label{totalaction}
\end{equation}
where $g_{\mu\nu}$ is the metric on $\a$ and $h_{ij}$ is the
induced $n$-metric on the boundary, i.e., $g_{{ij}|_{\partial
\a}}=h_{ij}$. $K$ is the trace of the extrinsic curvature $K_{ij}$
of the boundary defined according to the convention that the
outward normal to the boundary is positive.

The Einstein equation obtained from this action is
$R_{\mu\nu}=\frac{2\Lambda}{n-1}g_{\mu\nu}$. Recall that $\Lambda=
-\frac{n(n-1)}{2l^2}$ and that we set $l=1$. The
on-shell action therefore reads
\begin{equation}
I_{E} = -\frac{1}{16\pi G}\left(-2n\,\int_{\a}\sqrt{g}\, d^{n+1}x + 2\int_{\partial \a} \sqrt{h}\,d^n x \,
K\right).
\end{equation}
The first integral is the $(n+1)$-volume of $\a$ and the second integral
geometrically is the rate of change of the $n$-volume of the
boundary ${\partial \a}$ along the unit outward normal.

For a Schwarzschild-AdS$_{n+1}$ with an $S^{1}\times S^{n-1}$ boundary
at a constant radial distance it is fairly straightforward to
calculate both of above bulk and boundary contributions to the
on-shell action. They respectively are
\begin{equation}
V_{bh}=\int_{\a}\sqrt{g}\, d^{n+1}x=\frac{2\pi}{n\kappa}\left(\alpha^n-\rho_{_b}^n\right)\mathrm{Vol(S^{n-1})}
\end{equation}
and
\begin{equation}
\int_{\partial \a} d^n x
\sqrt{h}\,K=\frac{2\pi}{\kappa}\left(n\alpha^n+(n-1)\alpha^{n-2}-\frac{n}{2}M  \right)\mathrm{Vol(S^{n-1})}.
\end{equation}
Recalling that $M=\rho_{_b}^n+\rho_{_b}^{n-2}$, one obtains
\begin{equation}
I_{E_{bh}} = \frac{1}{4
G}\left(\frac{\rho_{_b}}{n\rho_{_b}^2+n-2}\right)\left((n-2)\rho_{_b}^n+n\rho_{_b}^{n-2}-2(n-1)(\alpha^n+\alpha^{n-2})\right)\mathrm{Vol(S^{n-1})}. \label{totact1}
\end{equation}
In the convention the $(n+1)$-dimensional Schwarzschild-AdS and AdS
metrics have been written, the base manifold $S^{n-1}$ (with the
canonical round metric on it) satisfies the $(n-1)$-dimensional
Einstein equation with a cosmological constant $(n-1)$. This is what
is referred to as the ``unit'' $(n-1)$-dimensional sphere in the
literature. Its volume
$\mathrm{Vol(S^{n-1})}=2\pi^{\frac{n}{2}}/\Gamma\left(\frac{n}{2}\right)$.

A few comments are in order. First, note that the action
(\ref{totact1}) is in terms of horizon radius. The actions for the two
infilling black hole solutions in terms of the boundary variables
$(\alpha,\beta)$ are obtained from it via substitution once the two
solutions of (\ref{eq}) are known. This takes us back to the classical
Dirichlet problem; explicit expressions for actions can be obtained
only when we know the explicit solutions of the infilling geometries.
Also note that action (\ref{totact1}) is finite for finite
$\alpha$. Hence, there is no {\emph{a priori}} need to do a background
subtraction (by taking off the action of AdS) in this case to make the
action finite as is needed in the infinite cavity limit \cite{HP}.
However, such a subtraction will appear in the next section in the
course of determining which classical solution will dominate the
path-integral for varying cavity radius and its wall temperature.

\subsection{Entropy and Specific Heats}
The canonical entropy of the system is
\begin{equation}
S=\beta \left(\frac{\partial I_{E_{bh}}}{\partial \beta}\right)-I_{E}
\end{equation}
in which the derivative is evaluated while keeping the area of
cavity fixed. The variation of $\beta$ changes $I_{E}$ through
variation of the mass, or equivalently the horizon radius
$\rho_{_b}$ of an infilling hole, and so
\begin{equation}
S=\beta \left(\frac{\partial I_{E_{bh}}}{\partial \rho_{_b}}\right)\left(\frac{\partial \rho_{_b}}{\partial \beta}\right) -I_{E}.
\end{equation}
One calculates the above two derivative terms using (\ref{totact1}) and
(\ref{betaasrho}). After some lengthy algebra one finds
\begin{equation}
S=\frac{1}{4G}\,\rho_{_b}^{n-1}
\end{equation}
precisely. This shows that the universal law of black-hole entropy
remains valid for such boundary conditions\footnote{It also confirms
that we have not missed out any factors in our action calculation
which will be important for doing precise calculations later in the
paper.}. This, to our knowledge, is the first derivation of this result
via the semi-classical path-integral approach.

It now immediately follows from the area-law and the study we made
in Section 3.4 on the variation of the mass of the two holes as functions
of wall-temperature that the specific heat
\begin{equation}
C_{A}=T\, \frac{\partial S}{\partial T}\label{ent}
\end{equation}
is positive for the larger black hole and negative for the smaller
black hole, and hence they are thermodynamically stable and unstable
respectively. The specific heat is zero when the two solutions are
degenerate.
\subsection{Phase transitions between hot AdS$_{n+1}$ and
Schwarzschild-AdS$_{n+1}$ within a finite cavity} To see that a phase
transition from hot AdS to Schwarzschild-AdS can occur within the
cavity, we need to compare the action of the infilling
Schwarzschild-AdS$_{n+1}$ solutions with that of hot AdS$_{n+1}$ space
which is unique for any given $S^1\times S^n$ boundary. We do so by
computing the actions of the infilling Schwarzschild-AdS$_{n+1}$
solutions in the ``background'' of the hot unique AdS$_{n+1}$ space,
i.e., by subtracting the action of the AdS$_{n+1}$ from that of the
black hole (\ref{totact1}). In the infinite limit of the cavity this
calculation simplifies as the boundary terms of the AdS and
Schwarzschild-AdS cancel \cite{HP,Witt1}. This simplification does not
exists for the finite cavity. Compared to the $\Lambda=0$ case, on the
other hand, the periodically identified AdS space has a non-trivial
Tolman redshift factor. This makes the boundary term more important
and the corresponding calculation more delicate than in flat space
\cite{AG}. One therefore needs to be cautious in subtracting the AdS
action from the Schwarzschild-AdS action in the case of finite
cavity. This is done as follows.  First calculate the volume and
boundary terms for AdS$_{n+1}$. They respectively are
\begin{equation}
V_{AdS}=\int_{\a}\sqrt{g}\, d^{n+1}x=\frac{\Delta t}{n}
\,\alpha^{n} \mathrm{Vol(S^{n-1})}
\end{equation}
and
\begin{equation}
\int_{\partial \a} d^n x
\sqrt{h}\,K_{0}=\Delta t\,\left(n\,\alpha^n+(n-1)\alpha^{n-2}\right)\mathrm{Vol(S^{n-1})} \label{bunads}
\end{equation}
where $\Delta t=2\pi\beta/(\sqrt{1+\alpha^2})$ as given by
Eq.(\ref{dirads00}).  Since the infilling hot AdS$_{n+1}$ space and
the two infilling black holes have the same $n$-metric on the
$S^1\times S^{n-1}$ boundary, we can replace $\beta$ using
Eq.(\ref{mast}), giving
\begin{equation}
\Delta t= \frac{2\pi}{\kappa}\sqrt{\frac{1-\frac{M}{\alpha^{n-2}}+\alpha^2}
 {1+\alpha^2}}
\end{equation}
where $M$ is either of the two infilling black hole masses. The reason
for replacing $\beta$ is to express the action in the same form as the
black-hole action. The volume and the boundary terms for AdS$_{n+1}$
then read
\begin{equation}
V_{AdS}=\int_{\a}\sqrt{g}\, d^{n+1}x=\frac{2\pi}{n\kappa}\sqrt{\frac{1-\frac{M}{\alpha^{n-2}}+\alpha^2}
 {1+\alpha^2}}
\alpha^{n} \mathrm{Vol(S^{n-1})},
\end{equation}
\begin{equation}
\int_{\partial \a} d^n x
\sqrt{h}\,K_{0}=\frac{2\pi}{\kappa}\sqrt{\frac{1-\frac{M}{\alpha^{n-2}}+\alpha^2}
 {1+\alpha^2}}\left(n\,\alpha^n+(n-1)\alpha^{n-2}\right)\mathrm{Vol(S^{n-1})}. \label{bunads}
\end{equation}
Note that because periodically identified AdS$_{n+1}$ space has a
non-trivial Tolman shift factor unlike hot flat space, the
boundary term (\ref{bunads}) contains contributions from the
changes in the volume of cavity as well as the local temperature
along the radial direction of the cavity, or equivalently, in the
Euclidean language: both changes in the $S^{n-1}$ base as well as
the radius of the $S^1$-fire at the $S^{1}\times S^{n-1}$ boundary
along the outward normal. The action of a black hole minus that of
hot AdS therefore reads
\begin{equation}
I_{E}=\frac{1}{4
G}\left(\frac{\rho_{_b}}{n\rho_{_b}^2+n-2}\right)\left((n-2)\rho^{n}_{_b}+n\rho^{n-2}_{_b}-2(n-1)(1-s)(\alpha^n+\alpha^{n-2})\right) \mathrm{Vol(S^{n-1})}\label{actionsubs}
\end{equation}
where, for short-hand,
\begin{equation}
s\equiv\sqrt{\frac{1-\frac{M}{\alpha^{n-2}}+\alpha^2} {1+\alpha^2}}\le
1.
\end{equation}
The action for either of the two infilling black holes is found by
substituting the corresponding solutions of $\rho_{_b}$ of
Eq.(\ref{eq}). The actions are then functions of cavity radius and the
wall temperature, i.e. expressed in terms of boundary data. Therefore
once analytic solutions of Eq.(\ref{eq}) are known the actions are
known exactly.
\subsubsection*{\it Phase Transition and the Critical Temperature}
The action (\ref{actionsubs}) is a smooth, single-valued function of
$\rho_{_b}$. It is zero for $\rho_{_b}=0$ and for $\rho_{_b}$ small
(compared to $\alpha$) the first two terms in the bracket dominate
and the action is positive. As one increases $\rho_{_b}$ it grows
monotonically to a maximum value and then decreases monotonically to
\begin{equation}
I_{E}(\alpha)=-\frac{1}{4G}\frac{\alpha}{n-2+n\alpha^2}\left(n\alpha^n+(n-2)\alpha^{n-2}\right)
2\pi^{\frac{n}{2}}/\Gamma\left(\frac{n}{2}\right)\label{endact}
\end{equation}
corresponding to $\rho_{_b}=\alpha$. This is negative definite. The
overall behaviour is similar to that in flat-space \cite{AG}. Below
(Figure 2-3) we plot $I_{E}$ for various values of $\alpha$ for
$n=3$. We choose four dimensions here because of its obvious physical
importance and also because the case of five dimensions, which is
physically interesting from a holographic point of view, will be
discussed in detail in the Section 5. We set $G=1$.
\begin{figure}[!h]
{\includegraphics[width=0.28\textwidth]{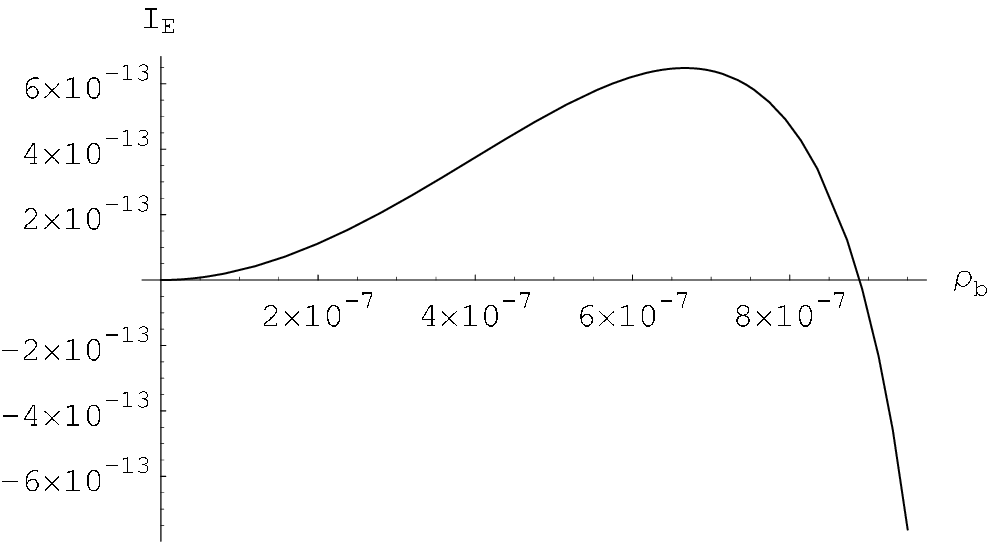}}
\hspace*{1.0cm}
{\includegraphics[width=0.28\textwidth]{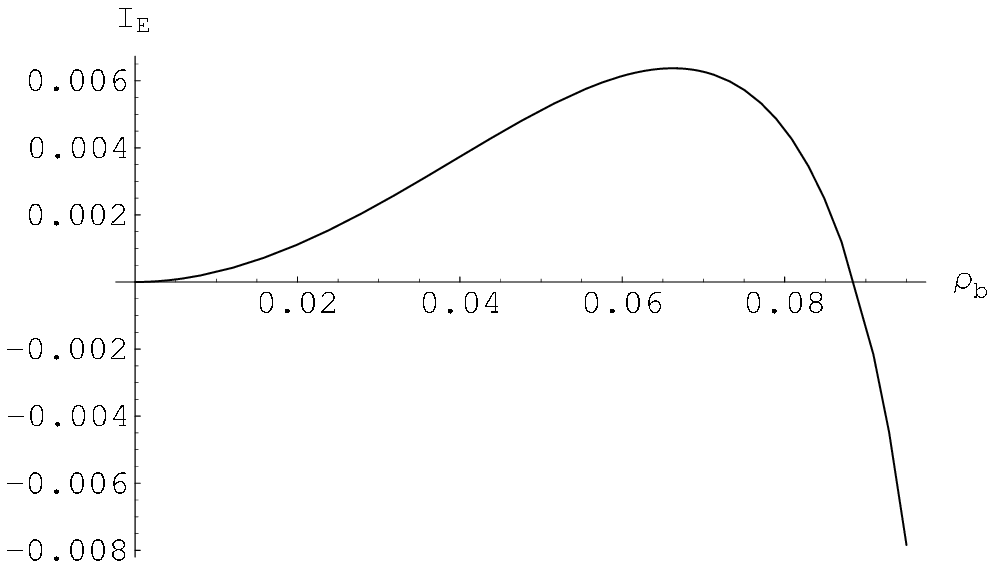}}
\hspace*{1.0cm}
{\includegraphics[width=0.28\textwidth]{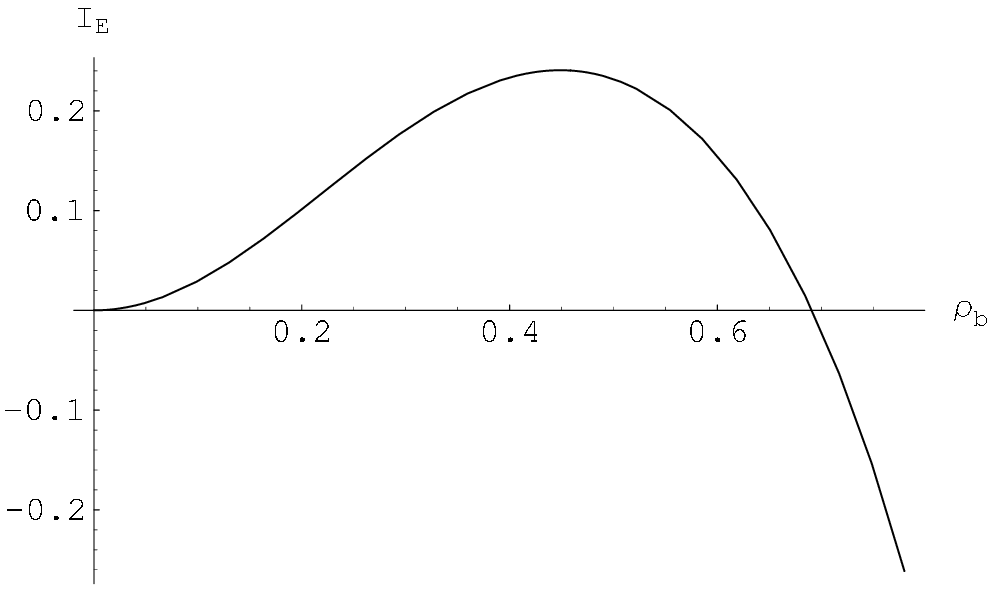}}
\caption{Actions in four dimensions as functions of $\rho_{_b}$
($\alpha=1\times 10^{-6},\,1\times 10^{-1},1.0$)}
\label{...}
\end{figure}
\begin{figure}[!h]
{\includegraphics[width=0.28\textwidth]{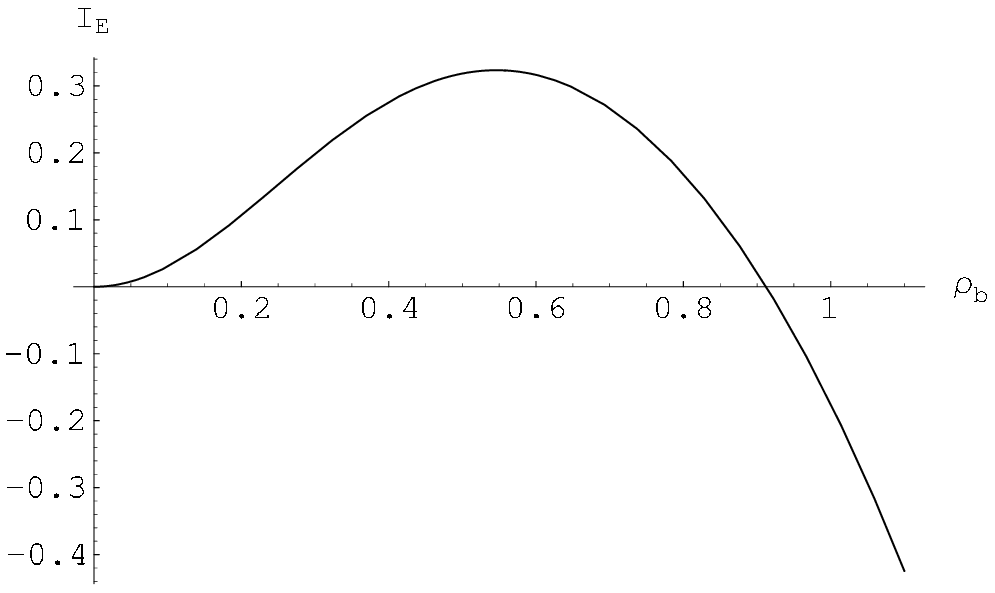}}
\hspace*{1.0cm}
{\includegraphics[width=0.28\textwidth]{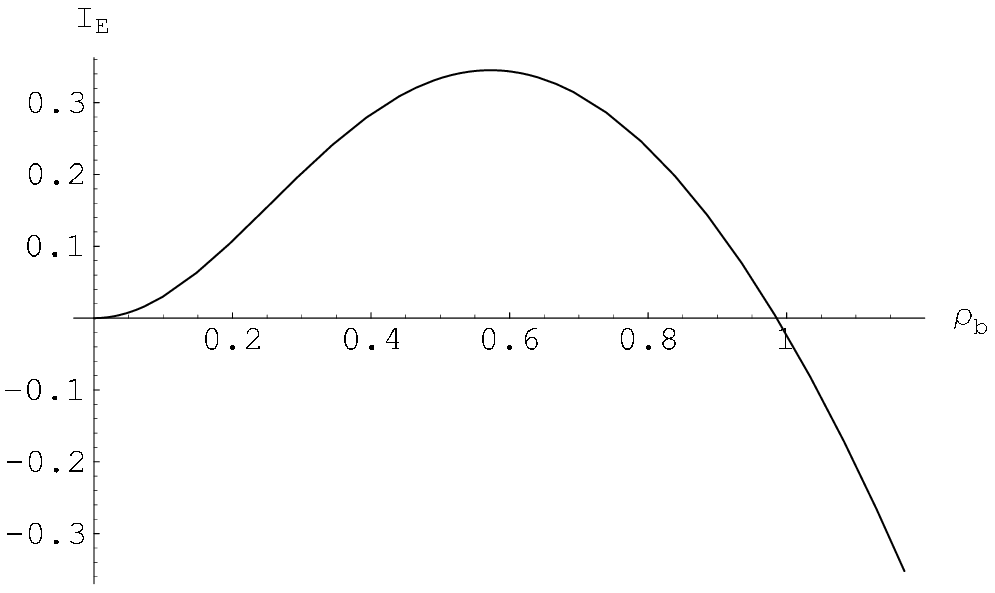}}
\hspace*{1.0cm}
{\includegraphics[width=0.28\textwidth]{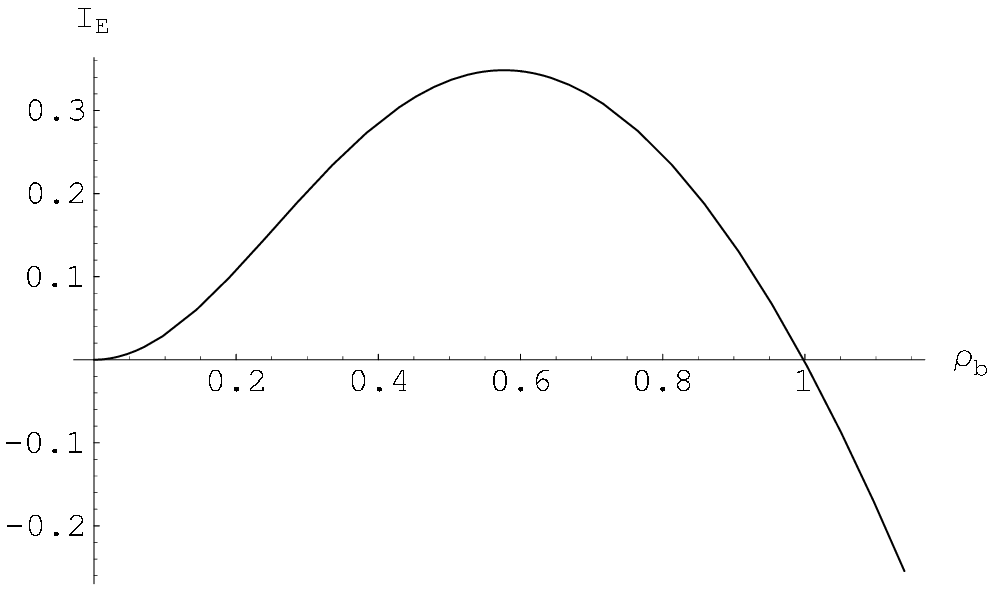}}
\caption{Actions in four dimensions as functions of $\rho_{_b}$ ($\alpha=2.0,\,4.0,\,8.0$)}
\label{...}
\end{figure}
This phase transition, however, is in terms of the horizon radius
(equivalently, the mass) of a black hole within the cavity. This shows
that within any finite cavity for sufficiently large value of its
horizon radius Schwarzschild-AdS becomes more probable than hot
AdS. This phase transition is possible for any value of the cavity
radius and is physical, i.e. the black hole does not engulf the
cavity. However, to revert to the canonical language we need to
understand how the phase transition takes place as function the
temperature of the wall instead of horizon-radius (or mass) of the
hole while the radius of the cavity is held fixed.  This needs care
and is not obvious because of the double-valuedness of the infilling
black-holes solutions. Can both of the two black-holes have larger
action than AdS? Does the scenario vary with varying radius of the
wall? To answer these questions we need to recall what we have learned
about the classical nature of the two solutions within the cavity. It
is not difficult to see that it is the larger mass solution that
nucleates from hot AdS for sufficiently high temperature of the
heat-bath. The action (\ref{actionsubs}) of the lower mass solution is
always positive thus making it the least probable solution
semi-classically.  To see this recall that there is an absolute upper
bound, $\rho_{_b}<\sqrt{(n-2)/n}$, of the horizon radius of the
smaller black hole. When the cavity radius is not too small the action
(\ref{actionsubs}) will always be positive within this range (Figure
3). On the other hand, when the cavity is small the action
(\ref{actionsubs}) (as well as the overall scenario) approaches the
flat-space limit (see Figure 2 above) and remains positive for the
smaller hole as discussed briefly above and in detail in
\cite{AG}. Therefore the action of the lower mass black hole is always
positive irrespective of the cavity radius and temperature. Only the
action of the larger mass hole is negative for $T>T_c$ and remains so
for all higher temperatures.

The critical temperature $T_c$ can be found by equating (\ref{endact})
to zero and eliminating the square roots.
In $(n+1)$ dimensions this
gives us an equation
\begin{equation}
(n-2)^{2}\rho_{_{bc}}^{n+2}+
2n(n-2)\rho_{_{bc}}^{n}+{n}^{2}{\rho_{_{bc}}}^{n-2}+
4(n-1){\alpha}^{n-2}(1+\alpha^2)\, \rho_{_{bc}}^{2} -4(n-1){\alpha}^{n-2}(1+\alpha^2)
=0 \label{mama}
\end{equation}
where $\rho_{_{bc}}$ is the radius of the larger black hole at the
critical temperature.

Once the solution of this equation is found, $T_c$ is determined via
(\ref{betaasrho}). Note that this equation has an appearance similar
to Eq.(\ref{minmin}). Using the identical arguments used for
Eq.(\ref{minmin}) it is easy to demonstrate that for large value of
the cavity-radius ${\rho_{_{bc}}}$ tends to unity. Also, this value is
the upper bound to $\rho_{_{bc}}$.  This is because like
Eq.(\ref{minmin}), the coefficients of various powers in (\ref{mama})
do not involve the cavity radius.  As in the case of $\Lambda=0$, the
horizon-radius of the nucleated black hole (or its mass) at the
critical temperature increases with the cavity radius. Therefore this
provides us with another example in which a sharp contrast exists
between the $\Lambda<0$ and $\Lambda=0$ cases. We will come back to
this issue in the next section when we discuss five dimensions
specifically and in the Conclusion.
\subsection{Infinite Cavity Limit}
For large $\alpha$
\begin{equation}
s=1-\frac{1}{2}\frac{m}{\alpha^n}+ \mathrm{higher\,\,\,order\,\,\,terms}.
\end{equation}
Therefore for $\alpha\rightarrow \infty$ one obtains the following action
of \cite{HP, Witt2}
\begin{equation}
I_{E}=\frac{1}{16\kappa G}\left(\rho_{_b}^{n-2}-\rho_{_b}^n\right)\mathrm{Vol(S^{n-1})}
\end{equation}
giving the Hawking-Page phase transition at $\rho_{_b}=1$. This
action is certainly much simpler than its finite boundary analogue
(\ref{actionsubs}).

It is straightforward to check the area-law of entropy and from it
the specific heats of the two solutions. Following our discussion
in Section 3.6 on the Hawking temperature it is easy to see that
the specific heat expression (\ref{ent})
\begin{equation}
C_{A}=T\, \frac{\partial S}{\partial T}\label{ent1}
\end{equation}
essentially reduces to one where $T$ can be replaced by $T_H$, the
Hawking temperature. Thus for any value of the Hawking temperature
(above the minimum $\sqrt{n(n-2)}/2\pi$, of course), the larger and
the smaller black hole solutions will have positive and negative
specific heats respectively. The critical (Hawking) temperature in
this case is $(n-2)/(2\pi)$, as one can check easily.
\section{Five Dimensions: Exact Results}
We have already remarked that Eq.(\ref{eq}) is a quintic in four
dimensions and the degree of this equation increases linearly with
dimensionality. However, an observation that we have not made so far
is that in odd dimensions Eq.(\ref{eq}) is an equation in
$\rho_{_b}^2$. For five and seven dimensions Eq.(\ref{eq}) is cubic
and quartic respectively and can be solved exactly using ordinary
algebraic methods. Thus the Dirichlet problem in these dimensions is
exactly solvable and one obtains the bulk geometries in terms of the
boundary variables. From a holographic perspective five dimensions is
special and we will treat it in detail. In this case Eq.(\ref{eq})
reads
\begin{equation}
z^{3}+\left(1+4{\alpha}^{2}{\beta}^{2
}\right)z^{2}+{\alpha}^{2}\left [4\,{\beta}^{2}-({
\alpha}^{2}+1)\right
]z+{\alpha}^{2}{\beta}^{2}=0 \label{eq5}.
\end{equation}
in which we have substituted $z$ for $\rho_{_b}^{2}$.

\subsection*{\it Explicit solutions}
The three roots of (\ref{eq5}) are given by Cardano's solution of the
cubic:
\begin{equation}
z_{1}=\frac{1}{6}(P+12
\sqrt{Q})^{\frac{1}{3}}-\frac{R}{6(P+12
\sqrt{Q})^{\frac{1}{3}}} -\frac{1}{3}(1+4\alpha^{2}\beta^{2}),
\end{equation}
\begin{equation}
z_{2}\!=\!\!-\frac{1}{12}(P+12
\sqrt{Q})^{\frac{1}{3}}\!-\frac{R}{12(P+12
\sqrt{Q})^{\frac{1}{3}}}
-\frac{1}{3}(1+4\alpha^{2}\beta^{2})-\frac{i\sqrt{3}}{2}\!\!\left(\!\!\frac{1}{6(P\!+\!12
\sqrt{Q})^{\!\frac{1}{3}}}\!-\!\frac{R}{(P\!+\!12
\sqrt{Q})^{\!\frac{1}{3}}}\!\!\right)
\end{equation}
and
\begin{equation}
z_{3}\!=\!\!-\frac{1}{12}(P+12
\sqrt{Q})^{\frac{1}{3}}\!-\frac{R}{12(P+12
\sqrt{Q})^{\frac{1}{3}}}
-\frac{1}{3}(1+4\alpha^{2}\beta^{2})+\frac{i\sqrt{3}}{2}\!\!\left(\!\!\frac{1}{6(P\!+\!12
\sqrt{Q})^{\!\frac{1}{3}}}\!-\!\frac{R}{(P\!+\!12
\sqrt{Q})^{\!\frac{1}{3}}}\!\!\right)
\end{equation}
where
\begin{equation}
P=-16\,{\alpha}^{6} {\beta}^{2}\left( 32\,{\beta}^{4} +9\right) +12\,{
\alpha}^{4} \left(16\,{\beta}^{4}-12\,{\beta}^{2} -3\right) -12\,{
\alpha}^{2} \left(5\,{\beta}^{2} +3\right) -8
\end{equation}
and
\begin{equation}
\begin{array}{rcl}
Q&=&-3\alpha^{2}\left (2{\alpha}^{2
}+1\right )^2 \\
&&\left ({\alpha}^{6}(4{\beta}^{4}+1)-\alpha^{4}(32{\beta}^{6}-4{\beta}^{4}+10{\beta}^
{2}-2)+{\alpha}^{2}(13{\beta}^{4}-10{\beta}^{2}+1)-4{\beta}^{2}\right
)
\end{array}
\end{equation}
and
\begin{equation}
R=4\alpha^{4}(16{\beta}^{4}+3)+4{\alpha}^{2}(3-4\beta^{2}).
\end{equation}
As usual with Cardano's solutions, the three roots of Eq.(\ref{eq5})
are given in an imaginary form and hence it is not obvious which two
are positive.  However, since we know that they will be
positive/complex together, it is easy to single out the expressions
for the two positive roots; they are $z_{1}$ and $z_{2}$ (see
below). These two solutions therefore give the masses of the two black
holes, and hence the infilling geometries in terms of the boundary
data.
\subsection{The minimum temperature $T_m$ for black holes}
In finding various other quantities of interest one still needs to be
judicious in choosing the correct algebraic approach as we will see
below. First we would like to know $T_m$ (above which the two black
hole solutions classically) as function of cavity radius. This has
been discussed qualitatively in Section 3.4 for arbitrary
dimensions. To find $T_m$ exactly, one is required to find the
positive root of (\ref{minmin}) and then obtain $T_m$ from
(\ref{betaasrho}). For five dimensions this is possible as
(\ref{minmin}) simplifies to a cubic equation like (\ref{eq5}).
However, the resulting algebraic expression for $T_m$ will not be
economical and easy to simplify.

A much nicer algebraic expression follows if one considers
Eq.(\ref{eq5}) directly and makes use of the known algebraic
methods. Eq.(\ref{eq5}) admits a negative root and pair of positive or
complex roots. These positive roots are in 1-1 correspondence with the
two positive roots of Eq.(\ref{eq}) (for $n=4$) and hence all
qualitative observations made earlier apply directly. In particular,
the positive roots of Eq.(\ref{eq5}) will appear and disappear
simultaneously together with the two positive roots of Eq.(\ref{eq})
as $\beta$ is varied. Because it is cubic and has a negative root, if
we ensure that Eq.(\ref{eq5}) has all real roots then two of them will
be positive automatically and will correspond to the two
Schwarzschild-AdS$_{5}$ infilling geometries. The condition for a
cubic equation to have all three roots real is well-known (see, for
example, \cite{AS}). In this case it reads
\begin{equation}
32\,{\alpha}^{4}{\beta}^{6}-{\alpha}^{2}\left(4{\alpha}^{4}+4\alpha^{2}+13\right)\,{\beta}^{4}+
2\left(5\,\alpha^{4}+5\,\alpha^{2}+2\right){\beta}^{2}-{\alpha}^{2}({\alpha}^{2}+1)^2\le0. \label{cond}
\end{equation}
When the equality of this equation holds, the two positive roots of
are degenerate. Therefore the maximum value of $\beta_m$ is simply the
solution of
\begin{equation}
32\,{\alpha}^{4}{\beta_m}^{6}-{\alpha}^{2}\left(4{\alpha}^{4}+4\alpha^{2}+13\right)\,{\beta_m}^{4}+
2\left(5\,\alpha^{4}+5\,\alpha^{2}+2\right){\beta_m}^{2}-{\alpha}^{2}({\alpha}^{2}+1)^2=0. \label{condeq}
\end{equation}
This is cubic in $\beta_m^{2}$ and admits a unique positive root for
any value of $\alpha$ giving
\begin{equation}
T_m=\frac{1}{2\pi}\sqrt{\frac{
A^{\frac{2}{3}}-840\alpha^{4}-856\alpha^{2}-215+16\alpha^{4}+32\alpha^{6}+13
A^{\frac{1}{3}}+4\alpha^{4}A^{\frac{1}{3}}+
4\alpha^{2}A^{\frac{1}{3}}}{96\,\alpha^{2}A^{\frac{1}{3}}}}. \label{cond2}
\end{equation}
where
\begin{equation}
A=64{\alpha}^{12}+192{\alpha}^{10}+8880{\alpha}^{8}+17440{\alpha}^{6}-10308{\alpha}^{4}-18996{
\alpha}^{2}-5291+192(2{\alpha
}^{2}+1)^{2}\sqrt {3\left ({\alpha}^{4}+{\alpha}^{2}+7\right )^{3}}.
\end{equation}
This is plotted in Figure 4. The smaller the cavity the higher is
$T_{m}$ in concordance with our observations in Section 3 (and Figure
1).
\begin{figure}[!htc]
  \begin{center}
    \leavevmode
    \vbox {
      \includegraphics[width=10cm,height=6cm]{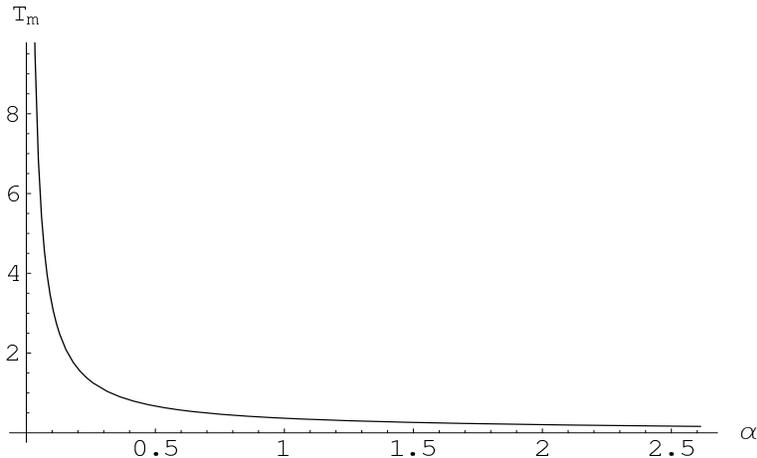}
      \vskip 3mm
      }
    \caption{The minimum temperature $T_m$ in five dimensions needed
    for the two black holes to exist
    classically. (For $T=T_m$ the solutions are degenerate.)}
    \label{Fig4}
  \end{center}
\end{figure}

\subsection{Actions, Phase transition and Critical Temperature}
The action (\ref{totact1}) in five dimensions is
\begin{equation}
I_{E} = \frac{1}{4G}\left(
\frac{\sqrt{z}}{2z+1}\right)\left(z^2+2\,z-3\left(\alpha^4+\alpha^{2}\right)\right)\mathrm{Vol(S^{3})}. \label{totact5}
\end{equation}
The actions of the two black holes are found by substituting $z_1$ and
$z_2$ in it. The action in the background of hot AdS space given by
(\ref{actionsubs}) reads
\begin{equation}
I_{E} = \frac{1}{4G}\left(
\frac{\sqrt{z}}{2z+1}\right)\left(z^2+2\,z-3(1-s)\left(\alpha^4+\alpha^{2}\right)\right)\mathrm{Vol(S^{3})}\label{5.12}
\end{equation}
in which
\begin{equation}
s=\sqrt{\frac{1-\frac{z^2+z}{\alpha^{2}}+\alpha^2} {1+\alpha^2}}.
\end{equation}
Again it is possible to find the actions of the two black holes
by substituting $z_1$ and $z_2$ directly in (\ref{5.12}).

\subsection*{\it Critical Temperature}
We have already shown that a phase transition takes the hot AdS 
to the larger mass Schwarzschild-AdS irrespective of the radius of the
cavity if the temperature of the cavity is above a certain critical
value. This critical value of temperature $T_c$ is a function of
cavity radius. We found that the radius of the black hole at the
critical temperature of a cavity does not increase indefinitely with
the cavity radius as in the case of vacuum. With the simplification
arising in the case of five dimensions we are able to treat these
issues exactly. First the radius of the black hole at the critical
temperature is obtained by equating (\ref{5.12}) to zero (or just
setting $n=4$ in (\ref{mama})):
\begin{equation}
{z}^{3}_c+4{z}^{2}_c+(3\alpha^4+3\alpha^2+4)z-3(\alpha^4+\alpha^2)=0. \label{positive}
\end{equation}
For any value of $\alpha$ this has one positive root. Again this is
cubic and hence can be solved exactly. The square-root of the positive
solution of Eq.(\ref{positive}) $\rho_{_{bc}}$ is plotted in Figure
5. We do not write the explicit form solution here to save space.
\begin{figure}[!htc]
  \begin{center}
    \leavevmode
    \vbox {
      \includegraphics[width=10cm,height=6cm]{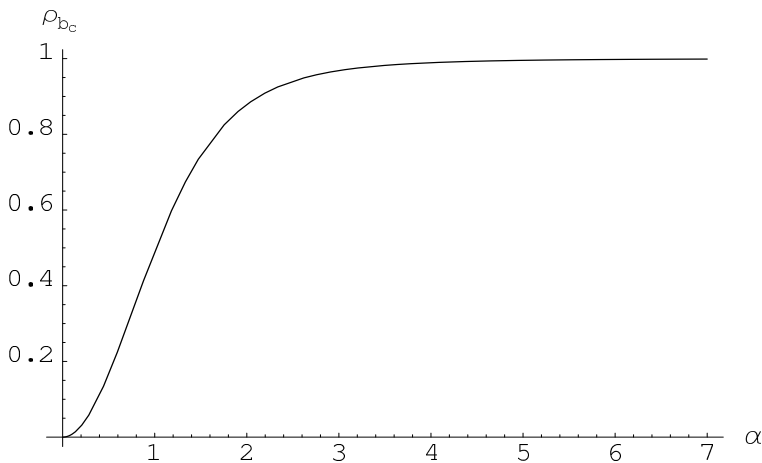}
      \vskip 3mm
      }
    \caption{The horizon radius of the black hole nucleated precisely
    at the critical temperature of the cavity as a function of cavity
    radius $\alpha$: its behaviour is different from that of
    $\Lambda=0$ case.}
    \label{Fig5}
  \end{center}
\end{figure}
The critical radius is thus found as an exact function of the cavity
radius and allows us to find the critical temperature $T_c$ needed for
the cavity to undergo a semi-classical phase transition.  It is
obtained simply by substituting the solutions of Eq.(\ref{positive})
in (\ref{betaasrho}) for $n=4$. Again, we do not mention the explicit
solution here as the form is not particularly illuminating. We plot
$T_c$ as a function of the cavity radius in Figure 6.
\begin{figure}[!htc]
  \begin{center} \leavevmode \vbox {
    \includegraphics[width=10cm,height=6cm]{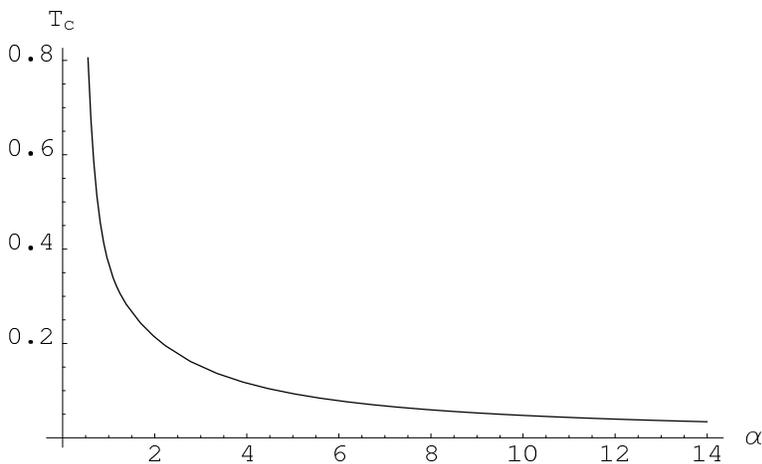} \vskip 3mm }
    \caption{The critical temperature $T_c$ increases with decreasing
    radius of the cavity as in $\Lambda=0$ case.}  \label{CritN}
    \end{center}
\end{figure}

\section{Conclusion}
In this paper we studied the thermodynamics of Schwarzschild black
holes in anti-de Sitter space within a cavity immersed in an
isothermal bath as functions of the cavity radius and its
wall-temperature, i.e., the temperature of the bath. This led us to
consider the associated Dirichlet boundary-value problem in Riemannian
geometry for an $S^1\times S^n$ boundary specified by its two radii
$(\beta, \alpha)$.  The connection between the Lorentzian and the
Euclidean pictures are such that the circumference of the $S^1$ fibre
gives the inverse temperature of the bath. Two topologically distinct
infillings are possible -- one with a nut (the Euclidean AdS) and the
other with a bolt (the Euclidean Schwarzschild-AdS). In the latter
case one finds that the infilling geometry is double-valued and exists
for a restricted range in the $\alpha-\beta$ plane\footnote{In the
case of $\Lambda<0$, for a boundary which is a non-trivial
$S^1$-bundle over $S^2$ the number of regular bolt-type infillings
(Taub-Bolt-AdS infillings) can be as high as ten \cite{Akbar:2003ed}
while the nut-type infillings (self-dual Taub-Nut-AdS infillings) are
unique \cite{Akbar:2002gv}, in the case of the latter explicit
solutions can be obtained. In the $\Lambda=0$ case, such boundaries
are studied for arbitrary dimensions in \cite{AG} where it is shown
that bolt-type infillings are always double-valued with the only
exception of Eguchi-Hanson metrics in which case the solutions are
unique as in the case of all nut-type infillings.}. We studied the
condition on $\alpha$ and $\beta$ for the existence of such classical
infillings and explored the possibility of obtaining their explicit
solutions in terms $\alpha$ and $\beta$ (or $T$).  We then considered
which solutions will dominate the path integral semi-classically as
one varies the boundary data by computing their actions and studying
them carefully. We found the overall qualitative picture to be similar
to that found in flat space \cite{York1, AG}: For any value of the
wall temperature a unique hot AdS solution exists whereas for
wall-temperatures below a minimum value $T_m$ no Schwarzschild-AdS
solutions exist.  Above this temperature there are two black-hole
solutions which continue to exist for all higher temperatures. As one
increases the temperature of the cavity the larger hole becomes
heavier and the smaller one gets lighter. We found the on-shell action
of the holes and showed that the standard area-law of black-hole
entropy holds for such boundary conditions from which one can
immediately deduce that the larger and the smaller holes have positive
and negative specific heats respectively as in the flat space case. At
$T_m$, both of the black holes have a positive action and hence the
AdS infilling is most probable and continues to remain so until a
critical temperature $T_c$ is reached when the action of the larger
hole is zero.  For $T>T_c$ the action of the larger hole becomes
negative definite. Note that as one increases the temperature of the
cavity from $T_m$, the mass of the smaller hole decreases and its
action decreases monotonically (and reaches zero at infinite
temperature).  But, although its mass increases monotonically with
temperature, the action of the larger hole increases only up to a
certain positive value before it starts decreasing monotonically to
zero at $T_c$ and becomes more negative for increasing temperatures
and reaches a fixed value (\ref{endact}) at infinite temperature. This
picture is the same for an arbitrary value of the cavity radius and
quantitatively approaches the flat space limit for small radii. Both
$T_m$ and $T_c$ decrease for larger value of the cavity radius as in
flat space.  However, despite the similarity, we have noted certain
differences between the $\Lambda<0$ and $\Lambda=0$ cases. As we have
seen in Section 3.4, there is an absolute upper bound on the horizon
radius of the smaller hole for $\Lambda<0$ whereas there is no such
absolute limit in the case of $\Lambda=0$ -- the upper limit in this
case is only on $x$ which translates to the limit (\ref{massFLAT}).
Another difference that occurs is in the horizon radius of the larger
hole nucleated at the critical temperature. This approaches an
absolute value of unity (in the units of the paper) for
$\Lambda<0$. In the flat space limit this gets larger and larger for
higher values of the cavity radius.

In the infinite limit of the cavity, on the other hand, the above
study reduces to the study made in \cite{HP}. We have shown how this
happens and how the only meaningful thermodynamic variable in this
limit is the Hawking temperature. In the Euclidean picture, this
corresponds to the fact that it is only the ratio of the radii of the
fibre and the base of the (infinite boundary) that is meaningful in
this limit-- this, in fact, is pivotal for the AdS/CFT correspondence
studied in \cite{Witt1}. The problem of finding explicit masses of
black hole solutions as functions of the Hawking temperature and
evaluation of other thermodynamic quantities simplifies enormously in
this limit.  Note that the phase transition found in this paper for
finite cavity does not follow from the Hawking-Page phase transition
by means of logical extrapolation. This is because the larger
black-hole solution and the hot AdS solution which induce the same
metric on the finite $S^1\times S^n$-boundary, do not have the same
$S^1\times S^n$-metric at infinity and if they match near infinity
they would not match anywhere at a finite radial distance. That there
is only one meaningful variable at infinity is slightly analogous to
the situation in the flat-space limit where, although both the radii
of the base and the fibre enter into the explicit solutions, it is
only the squashing of the two radii that is truly meaningful as one
can see from Eq.(\ref{mast0}). In the case of finite $S^1\times
S^n$-boundary both of the radii play a non-trivial role for
$\Lambda<0$.

All of the above results were obtained without taking recourse to
explicit infilling black-hole solutions for which one needs to solve
Eq.(\ref{eq}). This equation cannot be simplified by some redefinition
of variables and hence solutions are not possible for arbitrary
dimensions. However, the case of five dimensions turns out to be
rather special in which case this can be reduced to a cubic equation
and hence one can solve it exactly using ordinary algebraic methods
and thus obtaining the infilling black hole geometries as exact
functions of $\alpha$ and $\beta$ (or, $T$). This makes it possible to
compute the corresponding actions of the two black holes exactly as
functions of the boundary variables.  We have found $T_m$ and $T_c$ as
exact functions of cavity radius. The latter is an exact geometric
statement for any regular Euclidean Schwarzschild-AdS metric.  Other
quantities of interest can be computed from the action and the
solutions by using their standard definitions. These exact results
therefore provide a basis for further dynamical and thermodynamical
study and should find applications in brane-world cosmology,
holography and other related issues of current interest and are left
for future investigations.

\section*{Acknowledgements}
We would like to thank Gary Gibbons, Mahbub Majumdar and Danielle
Oriti for useful discussions. This work is funded by the Department of
Applied Mathematics and Theoretical Physics, University of Cambridge.


\begin{thebibliography}{99}
\bibitem{AS} M. ~Abramowitz and I.~A.~ Stegun (1965). \emph{Handbook of
Mathematical Functions} (Dover, New York).

\bibitem{Akbar:2003ed}
M.~M.~Akbar,
``Classical Boundary-value Problem in Riemannian Quantum Qravity and
Taub-Bolt-anti de Sitter Geometries,''
Nucl.\ Phys.\ B {\bf 663} (2003) 215
[arXiv:gr-qc/0301007].

\bibitem{Akbar:2002gv}
M.~M.~Akbar and P.~D.~D'Eath,
``Classical Boundary-value Problem in Riemannian Quantum Gravity and
Self-dual Taub-NUT-(anti)de Sitter Geometries,''
Nucl.\ Phys.\ B {\bf 648} (2003) 397
[arXiv:gr-qc/0202073].

\bibitem{AG}
M.~M.~Akbar and G.~W.~Gibbons,
``Ricci-flat Metrics with U(1) Action and the Dirichlet Boundary-value Problem in Riemannian Quantum Gravity and Isoperimetric Inequalities,''
arXiv:hep-th/0301026.

\bibitem{bailey}
W.~N.~Bailey (1935). \emph{Generalised Hypergeometric Series}
(Cambridge University Press, UK).

\bibitem{BC} S.~Barnard, and J.~M.~Child (1936). \emph{Higher Algebra}
(Macmillan, India).

\bibitem{BCM}
J.~D.~Brown, J.~Creighton and R.~B.~Mann,
``Temperature, Energy And Heat Capacity Of Asymptotically Anti-De Sitter Black Holes,''
Phys.\ Rev.\ D {\bf 50} (1994) 6394 [arXiv:gr-qc/9405007].

\bibitem{Birkeland} R.~Birkeland,
``\"{U}ber die Aufl\"{o}sung algebraischer Gleichungen durch
hypergeometrische Funktionen,''
Math. Zeitschrift 26 (1927) 565-578.

\bibitem{Braden}
H.~W.~Braden, B.~F.~Whiting and J.~W.~York,
``Density of States for the Gravitational Field in Black Hole Topologies,''
Phys.\ Rev.\ D {\bf 36} (1987) 3614.

\bibitem{GHsur}
G.~W.~Gibbons and S.~W.~Hawking,
``Action Integrals and Partition Functions in Quantum Gravity,''
Phys.\ Rev.\ D {\bf 15} (1977) 2752.

\bibitem{GH}
G.~W.~Gibbons and S.~W.~Hawking,
``Classification of Gravitational Instanton Symmetries,''
Commun.\ Math.\ Phys.\  {\bf 66} (1979) 291.

\bibitem{GR}
J.~P.~Gregory and S.~F.~Ross,
``Stability and the Negative Mode for Schwarzschild in a Finite Cavity,''
Phys.\ Rev.\ D {\bf 64} (2001) 124006
[arXiv:hep-th/0106220].

\bibitem{HP}
S.~W.~Hawking and D.~N.~Page,
``Thermodynamics of Black Holes in Anti-de Sitter Space,''
Commun.\ Math.\ Phys.\  {\bf 87} (1983) 577.

\bibitem{Maldacena}
J.~M.~Maldacena,
``The large N limit of superconformal field theories and supergravity,''
Adv.\ Theor.\ Math.\ Phys.\  {\bf 2}, 231 (1998)
[Int.\ J.\ Theor.\ Phys.\  {\bf 38}, 1113 (1999)]
[arXiv:hep-th/9711200].

\bibitem{Myers}
R.~C.~Myers and M.~J.~Perry,
``Black Holes in Higher Dimensional Space-Times,''
Annals Phys., NY\  {\bf 172} (1986) 304.

\bibitem{Nojiri}
S.~Nojiri, S.~D.~Odintsov and S.~Ogushi,
``Friedmann-Robertson-Walker Brane Cosmological Equations from the  Five-dimensional Bulk (A)dS Black Hole,''
Int.\ J.\ Mod.\ Phys.\ A {\bf 17} (2002) 4809
[arXiv:hep-th/0205187].

\bibitem{Tim}
T.~Prestidge, ``Dynamic and Thermodynamic Stability and
Negative Modes in Schwarzschild-anti-de Sitter,'' Phys.\ Rev.\ D {\bf
61} (2000) 084002 [arXiv:hep-th/9907163].

\bibitem{RS}
L.~Randall and R.~Sundrum,
``An Alternative to Compactification,''
Phys.\ Rev.\ Lett.\  {\bf 83} (1999) 4690
[arXiv:hep-th/9906064].

\bibitem{Regge1}
T.~Regge and C.~Teitelboim, ``Role of Surface Integrals in the
Hamiltonian Formulation of General Relativity,'' Annals Phys.\ {\bf
88} (1974) 286.

\bibitem{Regge2}
T.~Regge and C.~Teitelboim,
``Improved Hamiltonian for General Relativity,''
Phys.\ Lett.\ B {\bf 53} (1974) 101.

\bibitem{Stu}
B.~Sturmfels,
``Solving Algebraic Equations in Terms of ${\cal{A}}$-hypergeometric
Series,''
Discrete Mathematics 210 (2000) 171-181.

\bibitem{Witt1}
E.~Witten,
``Anti-de Sitter Space and Holography,''
Adv.\ Theor.\ Math.\ Phys.\  {\bf 2} (1998) 253
[arXiv:hep-th/9802150].
\bibitem{Witt2}
E.~Witten,
``Anti-de Sitter Space, Thermal Phase Transition, and Confinement in  Gauge Theories,''
Adv.\ Theor.\ Math.\ Phys.\  {\bf 2} (1998) 505
[arXiv:hep-th/9803131].

\bibitem{York1}
J.~W.~York,
``Black Hole Thermodynamics and the Euclidean Einstein Action,''
Phys.\ Rev.\ D {\bf 33} (1986) 2092.

\bibitem{York-sur}
J.~W.~York,
``Role of Conformal Three Geometry in the Dynamics of Gravitation,''
Phys.\ Rev.\ Lett.\  {\bf 28} (1972) 1082.




\end{thebibliography}
\end{document}